\documentclass[]{aa} 
\usepackage[dvips]{graphicx}
\usepackage{natbib}
\bibpunct{(}{)}{;}{a}{}{,}

\def\Msun{M$\sb{\odot}$}

\def\Teff{$T_{\rm eff}$}

\def\kms{km~s$^{-1}$}
\def\gtrsim{\,\,\raise0.14em\hbox{$>$}\kern-0.76em\lower0.28em\hbox  
{$\sim$}\,\,}  
\def\lesssim{\,\,\raise0.14em\hbox{$<$}\kern-0.76em\lower0.28em\hbox  
{$\sim$}\,\,}  

\begin{document}

\title{ More lead stars\thanks{Based on observations carried out at
the {\it European Southern Observatory} (La Silla, Chile; Program 65.L-0354) and
at the {\it Observatoire de Haute Provence} (operated by CNRS, France)}}

\author{
S. Van Eck\inst{1}\thanks{Post-doctoral Researcher, F.N.R.S., Belgium}
\and
S. Goriely\inst{1}\thanks{Research Associate, F.N.R.S., Belgium}
\and
A.~Jorissen\inst{1}$^{\star\star\star}$
\and
B. Plez\inst{2}} 

\institute{Institut d'Astronomie et d'Astrophysique, 
Universit\'e Libre de Bruxelles, C.P.~226, Boulevard du Triomphe, B-1050 
Bruxelles, Belgium
\and
GRAAL, Universit\'e de Montpellier II, cc 072,  F-34095 Montpellier Cedex 05, France
}
\date{Received date / Accepted date} 
\offprints{S. Van Eck \email{svaneck@astro.ulb.ac.be}}

\abstract{
The standard model for the operation of the s-process in asymptotic giant
branch (AGB) stars predicts that low-metallicity 
([Fe/H] $\lesssim -1$) AGB stars
should exhibit large overabundances of Pb and Bi as compared to
other s-elements. The discovery of the first three such `lead stars'
(defined as stars enriched 
in s-elements with [Pb/hs] $\ga 1$, hs being any of Ba, La or Ce) 
among CH stars
has been reported in a previous paper 
(Van Eck et al., Nature 2001, 412, 793). 
Five more CH stars (with [Fe/H] 
ranging from $-1.5$ to $-2.5$) are studied in the present paper, 
and two of them appear to be enriched in lead
(with [Pb/Ce]$\simeq 0.7$). 
The \ion{Pb}{I} line at $\lambda4057.812$~\AA\ is detected and clearly resolved thanks 
to high-resolution spectra ($R = \lambda/\Delta\lambda = 135\ts000$).
The abundances for these two stars (HD 198269 and HD 201626) 
are consistent with the predictions for the s-process operating in low-metallicity AGB stars 
as a consequence of the 
`partial mixing' of protons below the convective hydrogen envelope.
Another two stars (HD 189711 and V~Ari) add to a growing number of 
low-metallicity stars (also including LP 625-44 and LP 706-7, 
as reported by Aoki et al., 2001,
ApJ 561, 346) which do not conform to these predictions, however. 
Variations on the canonical proton-mixing scenario for the operation of the s-process in low-metallicity stars, that could account for these discrepant stars, 
are briefly discussed. 
}

\maketitle
%

\section{Introduction}

The overabundances of elements heavier than iron observed at the
surface of 
Asymptotic Giant Branch (AGB) stars \citep{Smith-Lambert-90}
clearly indicate that the s-process takes place during 
the AGB phase in the evolution of low- and intermediate-mass 
stars ($0.8 \le M (\mathrm{M}_{\odot}) \le 8~$). It is now widely accepted that
the neutron exposure required 
to produce s-elements originates in some  
partial mixing of protons (PMP) from the envelope down into the C-rich layers
resulting from the former intermittent operation of He-burning during 
thermal pulses.  
PMP activates the chain of reactions
$^{12}\mathrm{C(p,}\gamma)^{13}\mathrm{N}(\beta)
^{13}\mathrm{C}(\alpha,\mathrm{n})^{16}\mathrm{O}$. 
The s-elements thus produced in the deep interior by successive neutron captures
are subsequently brought 
to the surface by the third dredge-up (3DUP). 
The mixing scenario has recently received a new impetus 
from the inclusion in AGB models of mechanisms like
overshooting
\citep{Herwig-97} or rotation \citep{Langer-99} that trigger simultaneously
the 3DUP and the PMP.
Unfortunately, model predictions at the present time 
do not agree, neither on the mechanism responsible for the 3DUP or PMP, 
nor on their physical characteristics, such as their extent 
into the C-rich layers
or the number of pulses required for the 3DUP and PMP to set in 
\citep[see e.g.,][]{Lattanzio-98}.
It is therefore of prime importance to devise predictions that may be tested
against abundance observations. 
One such prediction is that  
stars with metallicities [Fe/H] $\le -1$
(where $[A/X]=$ log$(N_A/N_X) - $log$(N_A/N_X)_\odot$) 
should exhibit large overabundances of Pb--Bi as compared to 
lighter s-elements
\citep[Pb/Ba ratios as large as 70 are predicted in AGB stars with {[}Fe/H{]}$ = -1.3$;][]{Goriely-00}. Therefore, if the standard PMP scenario holds,
low-metallicity Pb-stars should exist. Such an observation would provide strong
support to the
state-of-the-art s-process models, provided that a possible r-process origin of
Pb be excluded from the absence of overabundances for r-process elements. 

The first three such lead stars (HD 187861, HD 196944 and HD 224959, with metallicities in the range 
$-2.0$ to $-2.5$) have been reported among CH stars 
by \citet[][hereafter Paper~I]{VanEck-01}. At the same time,  \citet{Aoki2001} found that the slightly more metal-deficient stars 
\object{LP 625-44} and \object{LP 706-7} are enriched in carbon and s-elements, and yet, with [Pb/Ce] $< 0.4$,
cannot be considered as lead
stars, in disagreement with the standard PMP predictions. 
To clarify this situation, it is important to derive heavy element abundances for more metal-deficient stars.
The present paper adds 5 more stars to this set.


\section{The stellar sample}

CH stars \citep{Keenan-42,Keenan-93} are good candidates 
to investigate the properties of the s-process at low metallicities, since 
these C-rich stars have very weak lines of the iron-group 
elements but enhanced lines of the heavy elements \citep{Vanture-92c}.

{\it Late-type} and {\it early-type} CH stars, that may be
distinguished from their spectral types {\it and}  
$^{12}$C/$^{13}$C ratios \citep[late-type CH stars have  
$^{12}$C/$^{13}$C $\sim 500$ as compared to $\sim
10$ for early-type CH stars;][]{Kipper-96, Aoki-97}, probably form two
distinct families, with possibly different evolutionary histories.
It has been suggested that the same dichotomy as that observed between intrinsic and
extrinsic S stars \citep{VanEck-Jorissen-00} may be present among CH
stars. In that framework, late-type CH stars would be identified with 
{\it intrinsic} stars (i.e., genuine low-metallicity AGB stars, as
confirmed by their relatively bright absolute visual magnitudes; see
Table~\ref{Tab:sample}), 
whereas early-type CH stars would be identified with 
{\it extrinsic} stars \citep[i.e., post-mass-transfer binaries, as confirmed 
by][]{McClure-Woodsworth-90}.

Late-type CH stars 
might thus be considered as the Population
II analogues of N-type carbon stars, and as such, are ideal targets for the present
purpose of studying the operation of the s-process at low
metallicity. However, not many are known,
so that our sample has been extended to early-type CH stars.
Being post-mass-transfer binaries, these stars may be
considered as the Population II analogues of barium stars. 
The mass-function distribution of these binaries is 
consistent with the invisible companion being a white dwarf. 
The envelope of the
giant star  has been polluted by C-rich matter transferred from the companion when it was
an AGB star. The chemical composition of early-type CH stars should therefore
bear the signature of the nucleosynthesis processes operating in low-metallicity
AGB stars, provided that chemical fractionation processes similar
to those operating in post-AGB stars \citep{VanWinckel-92, Waters-92}
did not
alter the composition of the accreted matter.

\begin{table*}
\caption[]{Basic parameters of the program stars. The mean
  heliocentric radial velocity has been derived from 9 to 13 spectral
  lines as clean as possible. Its standard deviation corresponds to
  the line-to-line scatter. For comparison, previous radial-velocity determinations 
from the literature are also listed. In the case where orbital
elements are available 
(`orbit' in column `Rem.'), the `literature' radial velocity has been
computed for the date of observation from the available ephemeris. 
}
\label{Tab:sample}
\renewcommand{\tabcolsep}{2pt}
{\small
\begin{tabular}{llllllllrcrllllll}
\hline
Name      & Sp. type & $V$    & $M_{\rm V}$       & $T_{\rm eff}$ & 
$\log g$      & C/O & Obs.  & S/N & JD & \multicolumn{2}{c}{Rad.
  vel. (km/s)} & Rem.\\
\cline{11-12}
          &          &        &                   &   (K)  &
              &     &       &     &(-2\ts451\ts000) & this work &literature\\                     
\hline
\smallskip\\
\noalign{\it Late-type  CH}
\smallskip\\
\object{HD 13826}  & C4,4 CH$^a$ & 8.52 & $-2.6^a$ & 3580$^{o,y}$
& $-0.2^y$       & 1.07$^w$;& ESO & 44 & 804.739 &
$-168.2\pm2.2$ & $-176.0^v$\\ 
= \object{V Ari}  &             &      &          &             
&                &  2$^o$ &     &    &         &
               &           \\ 
\object{HD 189711} & C4,3 CH$^a$ & 8.37 & $-2.7^n$ & 3500$^d$ & $0.5^d$ 
  & 1.26$^n$ & ESO & 27 & 805.541 & $-166.4\pm2.3$  & $-168.0^v$ \\
\smallskip\\
\noalign{\it Early-type  CH}
\smallskip\\
\object{HD 26}     & C0,0 CH$^a$ & 8.25 & +0.3$^a$         & 5170$^{x}$ 
& 2.2$^x$  & 1.1$^t$ & ESO & 86 & 804.828 & $-210.8\pm0.4$  &
$-212.9^v$&  binary?$^u$\\
\object{HD 187861} & C1p,1 CH$^a$ & $9.2$& $-1.3^a$ & 5320$^{x}$ 
& 2.4$^{x}$  & 8.0$^t$ & ESO  & 67 & 804.589 & $+8.6\pm0.7$  &
$-4.0^v$ & binary?$^d$\\ 
\object{HD 198269} & C1,0 CH$^a$ & 8.23 
&+0.73$^f$         & 4800$^{x}$ & 1.3$^x$  & 2.0$^t$& OHP & 117 &
779.438 & $-198.6\pm1.0$  & $-199.4^u$ & orbit$^u$\\
\object{HD 196944} & wk line GK &8.4  & $-1.5\pm0.5^m$  & 5250$^m$
 &1.7$^m$ & 0.87$^m$ & ESO & 140 & 804.479 & $-174.1\pm0.4$  & $-174\pm5^m$\\
                   & str. CH$^s$\\ 
\object{HD 201626} & C1,1 CH$^a$ & 8.13 &
$-0.5^a$; 0:$^c$  & 5190$^{x}$ & 2.25$^x$  & 4.0$^t$& OHP & 102 &
780.449 & $-149.4\pm0.8$  & $-149.4^u$ &  orbit$^u$ \\ 
\object{HD 224959} & C1,1 CH$^a$
& 9.5  &$-0.15^b$;$-0.5^a$& 5200$^{x}$ & 1.9$^x$  & 5.0$^t$& ESO & 38
& 804.546 & $-130.5\pm0.4$  & $-135.1^u$ & orbit$^u$\\

\hline\\
\end{tabular}
}
a. \citet{Sleivite-90} and references therein;
b. \citet{Eggen-72c};
c. \citet{Wallerstein-69};
d. This work (see Sect.~\ref{Sect:obs});
f. \citet{Wallerstein-Knapp-98}: Hipparcos parallax;
l. \citet{Vanture-Wallerstein-99}; 
m. \citet{Zacs-98};
n. \citet{Kipper-96};
o. \citet{Aoki-97};
q. \citet{Knapp-2000:b};
s. \citet{Bidelman-81};
t. \citet{Vanture-92b};
u. \citet{McClure-Woodsworth-90};
v. \citet{Duflot-1995};
w. \citet{Kipper-90};
x. \citet{Vanture-92a};
y. \citet{Ulla-97};
z. \citet{Luck-Bond-82}.
\end{table*}

The target stars for this paper and Paper~I  are listed in Table~\ref{Tab:sample}, along with the 
effective temperature and gravity adopted from the literature (except
for HD~189711; see Sect.~\ref{Sect:analysis}).

\section{Observations and data reduction}
\label{Sect:obs}

The present project was conducted during two separate observing runs.
A first run at the {\it Observatoire de Haute Provence} (August 21--23, 2000)
explored at medium resolution two
spectral windows containing Pb lines in the optical domain. A subsequent run at
the European Southern Observatory (September 16--17, 2000) then
secured 
high-resolution spectra of what appeared to be the most promising -- albeit blended
on the medium-resolution OHP spectra -- Pb line (\ion{Pb}{I}
$\lambda\ 405.781$~nm).

The OHP spectra were obtained on the {\it AURELIE} spectrograph
\citep{Gillet-94} mounted on the 1.52m-telescope equipped with a
2048$\times$1024 EEV CCD (13.5~$\mu$m pixels). 
Two different spectral settings were used. The
1200~rules/mm grating (with filter OG~515) delivering a resolution $R = 
\lambda/\Delta\lambda\ = $ 85\ts000
in the second order at 720.0~nm was used to explore the 722.897~nm \ion{Pb}{I}
line detected by \citet{Gonzalez-98} in FG~Sge. The 3000~rules/mm holographic
grating was then used to explore the 401.963, 405.781 and 
406.214~nm \ion{Pb}{I} lines with a resolution of 45\ts000.

The ESO spectra were obtained on the  
Coud\'e Echelle Spectrometer (CES) fed by the 
3.6\,m-telescope using the Very Long Camera (f/12.5) in the blue path,
the high-resolution image slicer and the thinned, back-side illuminated
CCD\#61 (EEV, 2K $\times$ 4K pixels).
A resolution of 135\,000 was achieved at the central wavelength
of 405.8~nm. The spectra approximately cover the wavelength range 404.5--407.1~nm.

The reduction of the CCD frames followed the usual steps (bias and flat-field
corrections, optimal extraction of the stellar spectrum, wavelength calibration)
which were performed within the `long' context of the MIDAS software package.  
The signal-to-noise (S/N) ratio (per pixel) achieved  in the
  extracted spectrum
 is listed in Table~\ref{Tab:sample} along with the  heliocentric
 radial velocities. Our radial velocities 
 agree with the orbital ephemeris for HD~198269 and HD~201626, but
 they differ by about 5~km~s$^{-1}$ for HD~224959, for an
 unknown reason. The newly determined velocity of HD~187861 strongly
 suggests that this star is a binary since the two available
 velocities differ by more than 12~km~s$^{-1}$. On the contrary,
 there is no clear evidence so far that HD~196944 is a binary
 star. The radial-velocity variations observed for V~Ari are not
 unusual for a SRb variable, and do not necessarily hint at the
 binary nature of V~Ari.

\section{Abundance analysis}
\label{Sect:analysis}

\begin{figure}[ht!]
\resizebox{\hsize}{!}{\includegraphics{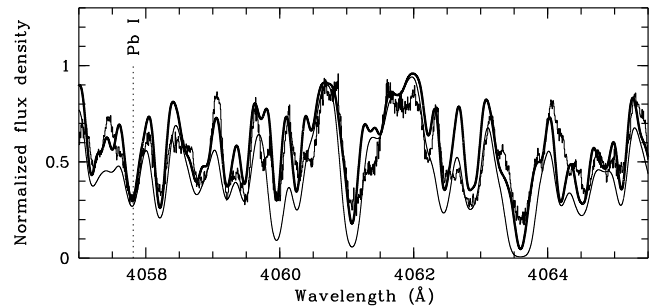}}
\caption[]{\label{Fig:microturb}
Impact on the synthetic spectrum of V~Ari of a change in the microturbulence velocity from
$\xi_t=2~$\kms\ (thick line) to $\xi_t=6~$\kms\ (thin line), all
abundances remaining the same. 
The intensity of the Pb~I $\lambda$ 405.78 nm line
(identified by a vertical dashed line) -- and hence the Pb abundance
-- is hardly affected by the change in microturbulence.
} 
\end{figure}

Abundances were derived using the {\it Turbospectrum} spectral synthesis
package  \citep{Alvarez-98}.
MARCS model atmospheres 
\nocite{Gustafsson-75,Plez-92b} 
(Gustafsson et al. 1975; Plez et al. 1992 and subsequent updates) 
matching the atmospheric parameters ($T_{\rm
eff} $ and $\log g$ as listed in Table~\ref{Tab:sample}), 
metallicity and C/O ratio have been used. A microturbulent velocity
$\xi = 2$~\kms\ has been adopted for all stars (the available
  spectral window is too narrow to allow a firm determination of the microturbulence).
Larger values for the microturbulence of  the two
late-type CH stars of our sample (HD~189711 and V~Ari) were adopted by 
\citet[][respectively  4~\kms\ and 6~\kms]{Tsuji-1991}.
Such large microturbulent velocities yield, however, 
a much worse fit of the observed spectrum than the smaller value of 2~\kms\ 
(see Fig.~\ref{Fig:microturb}). 
In any case, Fig.~\ref{Fig:microturb} shows that the intensity 
of the Pb~I $\lambda$ 405.78~nm line used in the abundance analysis (see below) 
is almost not altered by a change in the microturbulence velocity from 2 to
6~\kms. More precisely, if the value of the microturbulence velocity were to be 6~\kms,
the lead abundance would be diminished by less than 0.05 dex.

The C, N and O abundances play an important role in the strength of 
CN and CH lines present in our spectra. Since the C/O ratios taken 
from the literature  and listed in Table~\ref{Tab:sample} 
did not provide satisfactory
fits to the data, a better match was seeked by adjusting the C abundance on
the CN and CH lines present in the considered spectral window. 
Since the strength of CH lines is also sensitive to O through C$-$O (more than through C/O),
an independent estimate of the oxygen abundance is needed. It is
provided by the assumption that [O/Fe] = $+0.35$ in halo stars.
Since our C/O ratios are thus dependent  
upon this assumed O abundance, they  are not listed in
Table~\ref{Tab:sample} since they contain some degree of arbitrariness.
Estimates of the C, N and O abundances were in fact just needed to provide satisfactory  
fits to the molecular lines blending the atomic lines of interest.
Some mismatch persist, despite our efforts
to improve upon existing line lists for CN and CH \citep{lifbase}.

For the star HD~189711, no satisfactory fit whatsoever could be obtained for
$T_{\rm eff} = 3100$~K as adopted by \cite{Kipper-96}. It seems
necessary to adopt instead a somewhat larger temperature. 
The $B-V$ and $J-K$ indices of HD~189711
from \cite{Mendoza-1965:a} further support that conclusion, since
the observed indices  are not
consistent with the values derived from synthetic 
models at 3100~K, but rather point towards $T_{\rm eff} = 3500$~K, as
listed in Table~\ref{Tab:sample}. Note, however, that the gravity
listed Table~\ref{Tab:sample} is not well constrained.

\begin{figure}
\vspace{0cm}
\resizebox{\hsize}{!}{\includegraphics{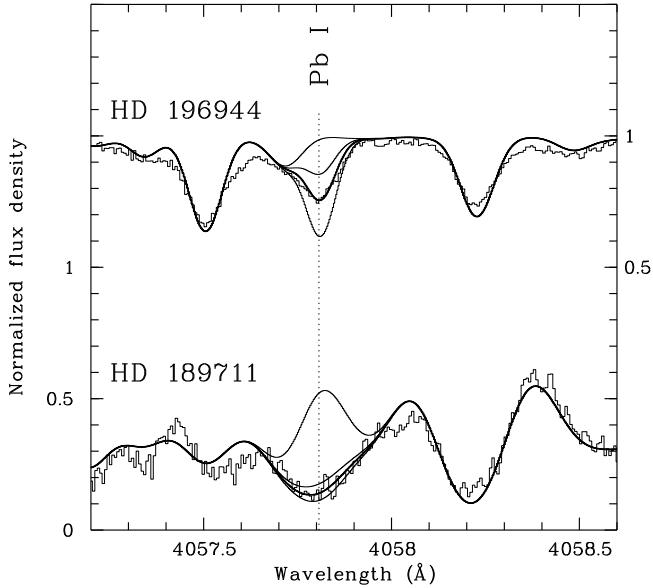}}
\caption[]{\label{Fig:Pb}
An illustration of the impact of changing the Pb abundance by
$\pm0.3$~dex on the intensity of the 405.781~nm 
Pb~I line for the two stars
HD 196944 ($T_{\rm eff} = 5250$~K) and  HD 189711 ($T_{\rm eff} = 3500$~K). 
For both stars, the thick line corresponds to the synthetic spectrum
obtained with the lead abundance
listed in Table~\ref{Tab:abundances} and the upper curve to no lead at all.
The scales on the left-hand and right-hand sides refer to HD~189711 
and HD~196944, respectively

} 
\end{figure}

Since the 406.214~nm and 722.897~nm \ion{Pb}{I} lines
are not visible on our spectra,
the Pb abundance was derived using the 405.781~nm Pb~I line only, 
including the contributions 
from the isotopes $^{204}$Pb, $^{206}$Pb, $^{207}$Pb, 
$^{208}$Pb as well as the hyperfine
structure due to $^{207}$Pb. The corresponding wavelengths and
$\log(gf)$ values are
listed in the Appendix.
To derive the Pb abundance in the presence of such an isotopic and
hyperfine structure requires to make an {\it a priori} assumption of
the isotopic ratios.
Nucleosynthesis calculations predict that the operation of the s-process in a low-mass,
low-metallicity AGB star results in highly non-solar isotopic ratios. 
In a $Z=6\;10^{-5}$ ([Fe/H] = $-2.5$), $M=0.9$~\Msun\ AGB star, 
$^{204}$Pb: $^{206}$Pb: $^{207}$Pb: $^{208}$Pb = 0.001: 0.026: 0.049: 0.924,
and these values have been adopted  in the present analysis.
Adopting the solar isotopic ratio instead ($^{204}$Pb: $^{206}$Pb:
$^{207}$Pb: $^{208}$Pb = 0.015: 0.236: 0.226: 0.523) typically decreases the Pb
abundance by about 0.1~dex, in agreement with the conclusions of
\citet{Aoki2002}. Fig.~\ref{Fig:Pb} presents the effect of changing
the Pb abundance by $\pm0.3$~dex on the intensity of the   405.781~nm
Pb~I line in different stars.

An attempt has been made to derive as well abundances for a few other elements 
(Fe, Zr, La, Ce, Nd and Sm) with useful lines falling in the
404.5 -- 407.1~nm spectral window. The oscillator strengths of these
lines have been taken from the VALD library
\citep{Kupka-99}. Whenever possible, these oscillator strengths have
been confirmed from a fit to the solar spectrum.
The comparison with the models presented in
Sect.~\ref{Sect:discussion} requires the Pb abundances to be
normalized by the abundance of an element of the hs group (Ba, La or Ce, where hs
stands for `heavy-s'), as well as an 
estimate of the metallicity. Despite the difficulty of this endeavour
on a spectral window covering only 3~nm in the violet spectrum of
carbon stars, 
such abundance ratios derived in a homogeneous manner (and thus less
prone to systematic errors) are probably preferable
over ratios obtained from La or Ce abundances
taken from the literature.
As it will be shown in
Sect.~\ref{Sect:discussion}, the distribution of low-metallicity,
s-process-rich stars in diagrams like ([Pb/La], [Fe/H])  and
([Pb/Ce], [Fe/H]) exhibits interesting properties, whose
identification is the main asset of the present paper.  
Of course, the abundance analysis will have to be redone on a firmer
basis when UVES/VLT spectra covering the full optical domain will become
available to us. A comparison with existing abundance data
(see below and Table~\ref{Tab:abundances}) makes us confident, 
however, that the
present abundance {\it ratios}\footnote{The abundance ratios listed in
  Table~\ref{Tab:abundances} are normalized by the meteoritic
  abundances of \citet{Grevesse-1998}, with $\log
  \epsilon({\rm Fe}) = 7.50$}, albeit preliminary, should not be too far off.    
As stressed above, the abundance ratios 
listed in Table~\ref{Tab:abundances}\footnote{The differences between the abundances obtained in this paper
and those from Paper~I for the three stars in common 
may be ascribed to the different metallicities
adopted in the two papers.} 
 must  be considered as tentative,
since, except for Ce, 
only few lines are available for each element, and they lie in a 
very crowded region of the spectrum, veiled by molecular lines. 
Since Ce has quite a few lines available, its abundance is probably more reliable than that
of the other heavy elements. 

The uncertainties  mentioned in Table~\ref{Tab:abundances} are estimated
from the range of abundances derived from the various available lines, 
although these uncertainties are not the standard deviation of the
abundances from the various lines, to account subjectively for the
different qualities of the different lines.
Errors caused by the uncertainties on the atmospheric parameters have
been estimated for HD~26 and V~Ari, and are listed in 
Table~\ref{Tab:errors} (see 
Paper~I for a similar error analysis in the case of HD 196944).

\begin{figure}[ht!]
\resizebox{\hsize}{!}{\includegraphics{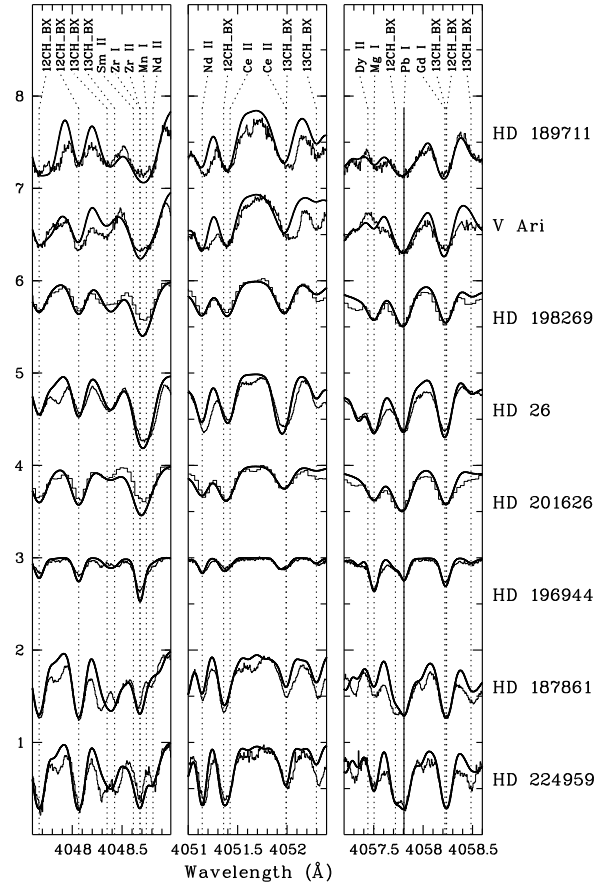}}
\caption[]{\label{Fig:fit}
A comparison between synthetic (thick lines) and observed
(thin lines) spectra. Ordinates are
flux densities normalized to the continuum.
Each spectrum is vertically shifted by one unit with respect
to the one just below.
The right panel includes the 405.78~nm Pb~I line
(indicated by a solid vertical line). The best fit is obtained for
HD~196944, the only non-carbon star in our sample, thus illustrating
the difficulties introduced by the presence of CN and CH lines in
this spectral region. The worst fits are obtained for the two coolest, 
late-type CH stars (HD 189711 and V~Ari)
} 
\end{figure}

The large uncertainties
on the abundances  for all the carbon stars (i.e., all stars
but HD~196944) are mainly due to
remaining inaccuracies in the CN line list. 
The general quality of the fit between the synthetic and observed 
spectra may be judged from Fig.~\ref{Fig:fit}.

Table~\ref{Tab:abundances} compares as well our abundances with those from the
literature. Our metallicities are systematically lower by 0.6 to 1~dex
as compared to those of \cite{Vanture-92c}, although the agreement
with other authors is better (see HD~196944 and V~Ari). It is
  worth noting in this respect that the metallicities derived in this
  paper are based on Fe lines falling in the 
Bond-Neff depression present in CH stars 
\citep{Bond-Neff-69}. This depression 
grossly covers the range 400--440~nm, and has a  
depth of about 10\% around 400~nm \citep{Gow-1976}. 
\citet{Luck-Bond-82} showed that the metallicity derived from Fe lines falling
in the Bond-Neff depression may be too low by as much as 1~dex.
This strengthens the importance of
deriving the Fe, hs and Pb abundances from the same spectral window,
and to discuss abundance ratios relative to Fe rather than relative to
H.
    
For the heavy elements, the agreement between the different
determinations appears to be much better, however, giving some
confidence to the abundances derived here. 
Despite the various difficulties encountered in the determination of
the abundances listed in Table~\ref{Tab:abundances} (narrow
wavelength range, contaminating CN and CH lines), the remarkable agreement
between our abundances  and
those obtained by \cite{Aoki2002} for HD 196944 is very encouraging.

\begin{table*}
\caption{\label{Tab:abundances}
Abundance ratios 
[X/Fe] for element X, normalized with the meteoritic abundances of
\cite{Grevesse-1998}, with $\log \epsilon({\rm Fe}) = 7.5$. 
The errors listed in the current table only
account for the uncertainties  in the line fitting and continuum placement (see
Table~\protect\ref{Tab:errors} for the error budget caused by the uncertainty
on the atmospheric parameters)
}
\begin{tabular}{l|ccccccccccccccccccccccccccccccccccccccccccc}
Star        & [Fe/H]   
            & [Zr/Fe]  
            & [La/Fe]  
            & [Ce/Fe]  
            & [Nd/Fe] 
            & [Sm/Fe] 
            & [Pb/Fe] & Ref.\\  
\hline
V Ari      & $-2.4 \pm 0.4$     
           & $ 1.1 \pm 0.25$    
           & $ 1.3 \pm0.5$       
           & 1.6${+0.5}\atop {-0.2}$ 
           & 1.9${+0.3}\atop{-0.4} $ 
           &         --          
           & 1.0${+0.2}\atop{-0.1}$ & (1)\\
           &      $-2.8$ 
                  &--
                  &--
                  &--
                  &--
                  &--
                  &--& (3)
\medskip\\
HD 224959   & $-2.2\pm 0.2$ 
            & $1.0\pm 0.1$  
            & $2.3\pm 0.2$  
            & $1.9\pm 0.3$  
            & $2.0\pm0.2$ 
            & $1.9\pm0.2$ 
            & $2.9\pm0.2$ & (1)\\
            & $-1.6$      
            & -- 
            & $2.0\pm0.6$ 
            & $2.1\pm0.1$ 
            & $1.8\pm0.3$ 
            & $1.4\pm0.15$  
            & -- & (2)
\medskip\\                
HD 26      & $-1.25\pm 0.3$ 
           & $ 0.9\pm 0.3$  
           &  2.3${+0.1}\atop{-0.5}$ 
           & 1.7 ${+0.3}\atop{-0.4}$ 
           &  1.3${+0.3}\atop{-0.1} $ 
           &  $1.3\pm 0.3$ 
           &  $1.9\pm 0.2 $ & (1)\\
           & $-0.4$ 
           & $0.9\pm0.3$ 
           & $1.4\pm0.3$ 
           & $1.9\pm0.4$ 
           & $1.6\pm0.15$ 
           & $1.9\pm0.2$  
           & -- & (2)\\
           &$-0.3$ 
           &-- 
           &-- 
           &-- 
           &-- 
           &-- 
           &--& (7)\\
           & $-0.45\pm 0.4$        
           & $0.60\pm 0.41$          
           & $2.27\pm0.34$
           & --
           & 1.99 
           & 0.40  
           & -- & (8)
\medskip\\ 
HD 187861  & $-$2.3${+0.5}\atop{-0.0}$  
           & $1.3 \pm 0.2 $ 
           & 2.2${+0.1}\atop{-0.15}$ 
           & $2.0\pm 0.3$ 
           & $1.9\pm0.1$  
           & $1.7\pm0.3$  
           & $3.1\pm0.1$   (1)\\
           &$-1.65$ 
           & -- 
           & $2.1\pm0.4$  
           & $2.0\pm0.2$ 
           & $1.8\pm0.2$ 
           & 0.7   
           & --       & (2)
\medskip\\
HD 196944 & $-2.4\pm 0.3$ 
          & $0.55\pm 0.1$  
          & $0.98\pm 0.1$  
          & $1.02\pm 0.1$ 
          & $0.7\pm 0.05$ 
          & $0.6\pm 0.2$  
          & $2.0\pm 0.05$ & (1)\\ 
          & $-2.45\pm0.1$ 
          & -- 
          & -- 
          & 1.5 
          & $0.9\pm0.2$  
          & --  
          & -- &  (6)\\
          & $-2.25\pm0.20$
          & $0.66\pm0.20$ 
          & $0.91\pm0.28$ 
          & $1.01\pm0.19$ 
          & $0.86\pm0.2$ 
          & $0.78\pm0.23$ 
          & $1.9\pm0.24$ & (9)
\medskip\\ 
HD 189711   & $-1.8 \pm 0.3$   
            &  1.0$\pm0.4$ 
            & $ 1.2\pm0.5$  
            &  1.7${+0.4}\atop{-0.6}$ 
            & 1.9$\pm0.4$ 
            & -- 
            &0.7${+0.5}\atop{-0.3}$ & (1)\\
            &$-1.15$ 
            & $0.95\pm0.5$ 
            & $1.95\pm0.5$ 
            & $2.35\pm0.5$
            & $2.1\pm0.5$ 
            & $1.8\pm0.5$ 
            & -- & (5)
\medskip\\
HD 198269    & $-2.2\pm 0.2$ 
             & $ 0.4\pm 0.1$ 
             & $ 1.6\pm 0.2$ 
             & $ 1.5\pm 0.3$ 
             & $1.2\pm0.1$ 
             & $1.2\pm0.2$ 
             & $2.2\pm 0.2$  & (1)\\
             & $-1.4$ 
             & $1.2\pm0.1$
             & $1.4\pm0.4$
             & $1.6\pm0.3$
             & $1.0\pm0.2$
             & $0.9\pm0.2$
             & --    & (2)
\medskip\\
HD 201626 & $-2.1\pm 0.1$
          & $ 0.9\pm 0.2$
          & $1.9 \pm 0.2$
          & $1.8 \pm 0.3$
          & $1.5\pm 0.2$
          & $1.4\pm 0.3$
          & $2.4\pm 0.1$ &(1)\\  
          & $-1.3$ 
          & $1.3$
          & $1.6\pm0.4$
          & $1.9\pm0.2$
          & $1.7\pm0.3$
          & $1.5\pm0.3$ 
          & --    & (2)
\medskip\\ 
\hline
\end{tabular}

References: (1) this work (2) \cite{Vanture-92c} (3) \cite{Kipper-90} (4)
\cite{Aoki-97} (5) \cite{Kipper-96} (6) \cite{Zacs-98} (7)
\cite{Thevenin-99} (8) \cite{Luck-Bond-82}
(9) \cite{Aoki2002}

\end{table*}

\begin{table*}
\caption{\label{Tab:errors}
Uncertainties on the derived abundances of HD~26 and V~Ari 
due to errors on the
atmospheric parameters (effective temperature \Teff, gravity $\log g$ and
microturbulence $\xi$). The values adopted for the reference model of
HD 26 are taken
from \citet{Vanture-92a}: \Teff\ = 5170~K, $\log g =2.2$, $\xi =
2.0$~\kms. For V~Ari, they are from \citet{Kipper-90}: 
\Teff\ = 3580~K and $\log g =-0.2$. 
The impact of the adopted  microturbulence ($\xi = 2$~\kms)
on the Pb abundance in
V~Ari has already been discussed in Fig.~\protect\ref{Fig:microturb}} 
 
\begin{tabular}{l|ccc|cc}
\hline
                 & \multicolumn{3}{c}{$\Delta \epsilon$ (HD 26)} &
                 \multicolumn{2}{c}{$\Delta \epsilon$ (V~Ari)} \\
\cline{2-4}\cline{5-6}
Element & $\Delta\xi = -0.5$ \kms &  $\Delta \log g = +0.3$ &  $\Delta T_{\rm
eff} = +100$ K &   $\Delta \log g = +0.3$ &  $\Delta T_{\rm
eff} = +100$ K \\ 
\hline
Fe    & +0.1     &  0.0       &  +0.1  &  0.0     & 0.0   \\
Zr    & +0.15    &  +0.1      &  +0.05 &  0.0     & +0.2  \\
La    & +0.1     &  +0.1      &  +0.05 &  0.0     & +0.0  \\
Ce    & +0.2     &  +0.1      &  +0.1  &  -0.05   & +0.1  \\
Nd    & +0.2     &  +0.1      &  +0.05 &  0.0     & 0.0   \\
Sm    & +0.1     &  +0.1      &  +0.1  &  0.0     & +0.1  \\
Pb    & +0.2     &  -0.05     &  +0.2  &  -0.1    & +0.2  \\
\hline 
\end{tabular}
\end{table*}

\section{Discussion}
\label{Sect:discussion}

Detailed heavy-element abundance patterns are available in the
literature for all the stars studied in this paper [Paper~I: HD
187861, HD 196944 and HD 224959; \citet{Vanture-92c}: HD 26, 
HD 187861, HD 198269, HD 201626 and HD 224959; \citet{Kipper-96}: HD
189711; \citet{Kipper-90}: V~Ari; \citet{Aoki2002}: HD~196944] and they 
reveal that it is the s- rather than the r-process which is
responsible for the observed overabundances. The situation is somewhat 
less clear for V~Ari, since \citet{Kipper-90} obtain [Eu/hs]
and [Gd/hs]$ \ga 0$, which cannot be accounted for by the
s-process. These authors obtain, however, extremely large overabundance
factors ([hs/Fe] of the order of 3 to 4 dex!), and caution the reader
about the large uncertainties affecting their results. 

\begin{figure}
\resizebox{\hsize}{!}{\includegraphics{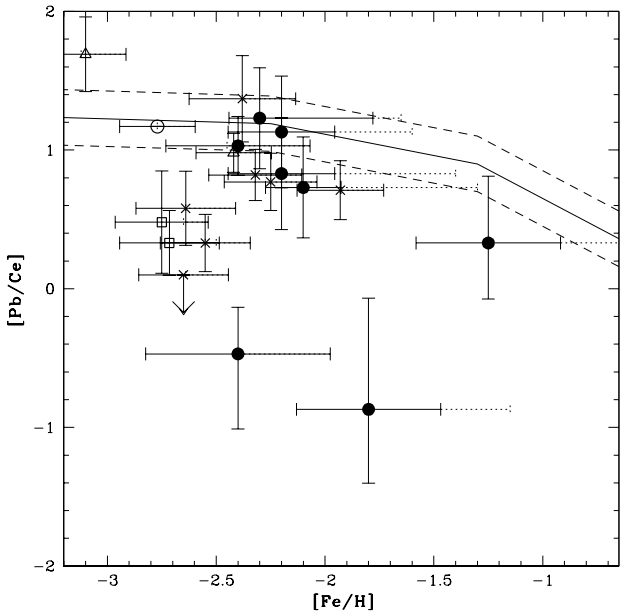}}
\caption[]{\label{Fig:PbCeFeH}
Evolution of the [Pb/Ce] ratio with
metallicity for the stars from this paper (filled circles), from
\citet[][open squares]{Aoki2001}, from \citet[][crosses]{Aoki2002},
from \citet[][open triangles]{Johnson-02} and
from \citet[][open circle]{Sivarani-02}.
The solid error bar is the root sum square of the errors associated
with the spectral fit, as  listed in
Table~\protect\ref{Tab:abundances}, and with the uncertainties on the
model parameters, as listed in Table~\ref{Tab:errors}.
The dotted error bar
includes other metallicity determinations from the literature.
Predictions from the standard PMP model from
\citet{Goriely-00} and \citet{Goriely-Siess-01} 
are represented by the solid line. The dashed
lines provide a rough estimate of the uncertainties on these
predictions, originating from the 
unknown dilution factors (i.e., number of pulses, and extent of PMP and
3DUP) and from the arbitrary proton mixing profile. } 
\end{figure}

Fig.~\ref{Fig:PbCeFeH} reveals that
HD~187861, HD 196944, HD 198269, HD 201626 and  HD 224959 all belong to the
class of `lead stars', since their
[Pb/Ce] ratios comply with the predictions for the standard model for 
PMP operating in low-metallicity AGB stars. 
  The two stars HD~189711 and V~Ari deviate from the expected
trend, however, since they have Pb/Ce ratios far too small with
respect to these predictions (the large uncertainty on the metallicity 
of HD~26 makes the situation unclear for that star). HD~189711 and
V~Ari are the two coolest CH stars studied in the present paper, and
abundance determinations in cool carbon stars represent a real
challenge to the spectroscopist. Nevertheless,
these two late-type CH stars add to similar (albeit much warmer) cases 
of metal-poor, s-process-rich, {\it non-lead} stars 
uncovered by \citet{Aoki2001,Aoki2002}.
All low-metallicity, s-process-rich  stars studied so far \citep[Paper~I;
this paper;][]{Aoki2001, Aoki2002, Johnson-02} 
are displayed in the ([Pb/Ce],[Fe/H]) 
diagram displayed in 
Fig.~\ref{Fig:PbCeFeH}. Clearly, not all low-metallicity,
  s-process-rich  stars
comply with the predictions of the standard PMP theory. 
The large uncertainty on the metallicity (see the
discussion in Sect.~\ref{Sect:analysis} and Fig.~\ref{Fig:PbCeFeH}),
does not alter this
conclusion, especially since the [Pb/Ce] does not appear to be very sensitive to
the adopted metallicity. This can be judged from the [Pb/Ce] ratios
obtained in this paper and in Paper~I for two different choices of the 
metallicity for the stars HD 187861 and HD 224959. Decreasing the
metallicity by about 0.5~dex resulted in a decrease of the [Pb/Ce]
ratio of only 0.15~dex.  

\begin{figure}
\resizebox{\hsize}{!}{\includegraphics{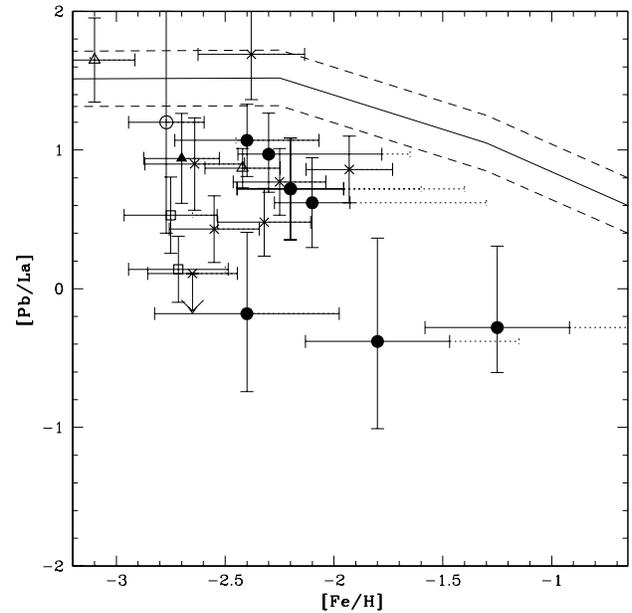}}
\caption[]{\label{Fig:PbLaFeH}
Same as Fig.~\protect\ref{Fig:PbCeFeH} for the [Pb/La] abundance ratio
}
\end{figure}

Some low-metallicity,
  s-process-rich  stars deviate as well from the standard PMP
  predictions in  the ([Pb/La], [Fe/H])
diagram (Fig.~\ref{Fig:PbLaFeH}), indicating that this behaviour is
not dependent upon the normalizing element (Ce or La). The agreement
with the model predictions is, however, not as good for [Pb/La] as it
is for [Pb/Ce], the observed  La abundances being systematically larger 
than the predicted ones.

The scatter in [Pb/Ce] and [Pb/La] at a given metallicity
  observed in Figs.~\ref{Fig:PbCeFeH} and \ref{Fig:PbLaFeH} is
  reminiscent of the situation prevailing among the (intrinsic, non-binary) post-AGB stars  
studied by \citet{VanWinckel-00}, who found a large scatter of [hs/ls]
also for s-process-rich stars with 
metallicities typical of the disk ($-1 \le {\rm [Fe/H]} \le -0.3$).

It should be stressed here that the predictions of the standard PMP
scenario are rather robust 
\citep[see the discussion by][]{Goriely-00}, in the sense that they 
do not depend upon the
details of the diffusive mixing (like its depth) as long as the proton profile
assumes a monotonic evolution with depth. In the framework of the
standard PMP scenario, there is thus no obvious degree of freedom that
could be used to reduce the lead production in low-metallicity AGB
stars, as seems to be required by the results presented in
Figs.~\ref{Fig:PbCeFeH} and \ref{Fig:PbLaFeH}.
The situation may change when AGB models will include the effect of
rotation, but such models are currently very preliminary
and do not include s-process calculations \citep{Langer-Herwig-01}. 
 We do not
endorse therefore the claim by
\cite{Aoki2001} that the Pb abundance observed in LP 625-44 could be
accounted for by reducing the $^{13}$C abundance in the $^{13}$C pocket
by a factor 20, since there is no freedom to do so in the framework of
the PMP scenario, where the $^{13}$C profile is fixed by the proton
profile. The $^{13}$C pocket required for the operation of the s-process
will form at that location in the intershell layer where the proton to
$^{12}$C ratio is just slightly below unity.  Above unity, the CN chain
would set in, producing a large amount of the $^{14}$N neutron poison;
too far below unity, the $^{13}$C abundance will be very low and its
contribution to the s-process will be largely overidden by the
contribution from the upper layers where
$^{13}$C is far more abundant. Such a layer with p/$^{12}$C $\la 1$
necessarily exists in the framework of the PMP scenario, since 
p/$^{12}$C $ > 100$ in the
outer hydrogen envelope, and goes down to 0 at the bottom  of the proton
diffusion layer. Finally, the
$^{13}$C abundance in the layer where p/$^{12}$C $\la 1$ is fixed by the $^{12}$C
abundance 
\citep[see for instance Eq.~8 in the analytical description provided
by][]{Jorissen-Arnould-89}, 
which is in turn fixed by the
properties of He-burning in the thermal pulse, leading inevitably to
$X(^{12}$C) $\sim 0.2$ to 0.5. On the other hand, if no dredge-up
occurs, PMP takes place in He-rich layers depleted in $^{12}$C
(typically  $X(^{12}$C) $\sim 10^{-4}$) and the $^{13}$C production is
too small to trigger any s-process.

A scenario to account for the low-metallicity,
  [Pb/hs]$\approx0$ stars has been promoted by \citet{Iwamoto-03}
and invokes the sudden mixing of protons from
the envelope directly into the thermal pulse. 
Detailed simulations remain, however, to be
performed in that framework.

For some of the stars from Fig.~\ref{Fig:PbCeFeH}, 
the comparison of the observed
abundances   with predictions from AGB
nucleosynthesis may even be questionable since some of them are
subgiant stars with no evidence for binarity \citep[LP
706-7;][]{Norris-97}. Their s-process enrichment can therefore not be 
ascribed to pollution by mass transfer from a former AGB companion. 
To make the situation even more
obscure, this conclusion {\it cannot} be generalized to {\it all} the
non-lead stars in Fig.~\ref{Fig:PbCeFeH}, since some are indeed binaries  
\citep[LP 625-44, CS 22942-019;][]{Preston-Sneden-01, Aoki2000, Aoki2002}, and might be relevant
to the mass-transfer scenario. 
Similarly, lead stars include {\it bona fide}
post-mass-transfer binaries \citep[like CS 29526-110 and the classical CH binaries HD
198269, HD 224959;][]{McClure-Woodsworth-90,
Aoki2002} as well as presumably single stars \citep[CS
22898-027, CS 22880-074;][]{Preston-Sneden-01, Aoki2002}. 
A whole new class of nucleosynthesis processes must be at work in the
non-binary  stars, or at least, their abundance peculiarities must have
their origin in a very tricky scenario. One such scenario has
recently been promoted by \citet{Thoul02}. It assumes that these
peculiar stars have evaporated from a globular cluster, where they
acquired their s-process enrichment from the pollution by the wind from a
nearby AGB star. 

\section{Conclusion}

We presented in this paper lead abundances for  8 low-metallicity,
s-process-rich stars. Although 5 among these conform with the predictions of
the partial proton-mixing operating in low-metallicity AGB stars (and
may therefore be tagged `lead stars'), two do not
since their [Pb/La]  and [Pb/Ce] ratios are too low by at least one order of
magnitude with respect to the predictions.  These two stars add to
similar cases uncovered by \cite{Aoki2001,Aoki2002}.
A large scatter is observed 
in the ([Pb/hs],[Fe/H]) diagrams collecting  
all low-metallicity, s-process-rich stars studied so far.
Lead stars
conforming to the [Pb/hs] predictions from the standard PMP scenario
co-exist with 
stars having [Pb/hs] ratios 
up to 2~dex lower. The existence of non-lead stars at low
metallicities is especially puzzling, since 
no physical phenomena can at the present time be called for
to explain, within the PMP scenario, the observed abundances.

Moreover, to add to the puzzle, some of the non-lead stars (especially
LP~706-7 and CS~22880-027) are subgiant stars with no evidence for
binarity, so that their s-process enrichment seems to call
for a rather exotic scenario (new site for the operation of the
s-process, or accretion from an s-process-rich wind from a nearby AGB
star in a globular cluster and subsequent evaporation thereof).

\acknowledgements
We would like to recognize G. Bihain for help with the 
abundance analysis of HD 26.

\appendix 

\section{Hyperfine and isotopic shifts of the Pb~I $\lambda$ 405.78 nm line}

The Pb~I $\lambda$ 405.78~nm line has a complex structure consisting
of three components arising from the hyperfine structure due to 
$^{207}$Pb plus three isotopic components associated with   
$^{204}$Pb,  $^{206}$Pb and $^{208}$Pb.
We present below a derivation of the position and intensities of these 
various components, which differ from the line list obtained by 
\cite{Aoki2001}

\subsection{Energy levels for pure $^{208}$Pb}

The energy levels for the  $^{208}$Pb isotope 
have been taken from \cite{Wood-68}, and yield 
a difference in energy of 24636.8973 cm$^{-1} (\pm 0.0012)$ for the
405.78~nm transition ($^3P_1 - ^3P_2$).

\subsection{Isotopic shifts}

Isotopic shifts for the 405.78~nm transition were taken from Table~I of
\cite{Bouazza00}.

\subsection{Hyperfine structure of $^{207}$Pb}

\begin{figure}
\resizebox{\hsize}{!}{\includegraphics{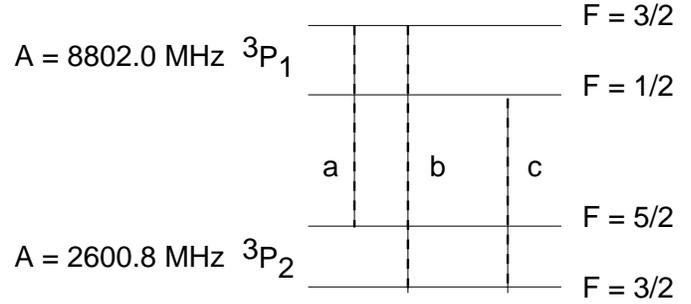}}
\caption[]{\label{Fig:hyperfine}
Energy levels and transition assignments for the hyperfine
structure of the Pb~I 405.78~nm line  
}
\end{figure}

The hyperfine structure of the levels involved in the 405.78 nm
transition are presented in Fig.~\ref{Fig:hyperfine}. 
The Russel-Saunders coupling of
the total electronic angular momentum $J$ (orbital + spin) with the
nuclear spin momentum ($I = 1/2$) predicts three components $a, b$ and 
$c$, with respective intensity  weights  of 9/15, 1/15 and 5/15, in agreement
with the experimental results of \cite{Manning50}.

The energy shifts are predicted by the Casimir formula,
using the magnetic hyperfine structure constant $A$ provided by 
\cite{Bouazza00}    as listed in  Fig.~\ref{Fig:hyperfine} 
($B = 0$, since $I = 1/2$).

The isotopic shift taken from Table I of \cite{Bouazza00} has then been added
to the hyperfine shift, to yield the final energy levels listed in
Table~\ref{Tab:linelist}.

\subsection{Wavelengths and  oscillator strengths for the Pb I $\lambda
405.78$~nm transition}

The final results are listed in Table~\ref{Tab:linelist}. 
The line structure is consistent with  that observed in the laboratory by
\citet[][his Fig.~2a]{Simons-89}, but not with the line list provided by
\cite{Aoki2001}. The $\log gf$ listed in Table~\ref{Tab:linelist} is that from
\citet[][$\log gf = -0.22$]{Biemont-2000},  weighted  by the hyperfine intensity
ratios in the case of $^{207}$Pb.

\begin{table}

\caption[]{\label{Tab:linelist}
Wavenumbers $\sigma$, air wavelengths $\lambda$  and  oscillator strengths
$f$ for the Pb I $\lambda 405.78$~nm transition}
\begin{tabular}{llll}
\hline
isotope/transition & $\sigma$& $\lambda$ & log(gf) \\
                                   & (cm$^{-1})$ & (\AA) \\
\hline
$^{208}$Pb   & 24636.8973 & 4057.807 & -0.220 \\
$^{207}$Pb a & 24636.9050 & 4057.805 & -0.442 \\
$^{207}$Pb b & 24637.1219 & 4057.770 & -1.396 \\
$^{207}$Pb c & 24636.6815 & 4057.842 & -0.697 \\
$^{206}$Pb   & 24636.8163 & 4057.820 & -0.220 \\
$^{204}$Pb   & 24636.7443 & 4057.832 & -0.220 \\
\hline
\end{tabular}
\end{table}

\bibliographystyle{apj}
  
\bibliography{/home/bibtex/ajorisse_articles,/home/bibtex/svaneck_books}

\end{document}